\documentclass{article}
\usepackage{spconf,amsmath,graphicx}

\renewenvironment{thebibliography}[1]{
  \begin{oldthebibliography}{#1}
    \setlength{\itemsep}{0.25em}
    \setlength{\parskip}{-0.35em}
}
{
  \end{oldthebibliography}
}

\usepackage{enumitem}
\usepackage{amsmath}
\usepackage{amssymb}
\usepackage{tabularx}
\usepackage{booktabs}
\usepackage{algorithm}
\usepackage{algpseudocode}
\usepackage{graphicx}
\usepackage[labelfont=bf]{caption}
\usepackage{float}
\usepackage{xcolor}
\usepackage[export]{adjustbox}

\setlist{nosep, leftmargin=14pt}

\usepackage{mwe} 

\title{Uncertainty-Aware Test-Time Adaptation for Inverse Consistent Diffeomorphic Lung Image Registration}
\makeatletter
\def\@name{{\em Muhammad F. A. Chaudhary}$^{\star}$\quad {\em Stephanie M. Aguilera}$^{\star}$\quad {\em Arie Nakhmani}$^{\star}$\quad {\em Joseph M. Reinhardt}$^{\dagger}$\\*[3pt]
  {\em Surya P. Bhatt}$^{\star}$\quad {\em Sandeep Bodduluri}$^{\star}$\\}
\makeatother

\address{$^{\star}$ Center for Lung Analytics and Imaging Research (CLAIR),\\ The University of Alabama at Birmingham, Birmingham, AL, USA \\
    $^{\dagger}$The Roy J. Carver Department of Biomedical Engineering, The University of Iowa, IA, USA}
\begin{document}
%
\maketitle
\begin{abstract}
Diffeomorphic deformable image registration ensures smooth invertible transformations across inspiratory and expiratory chest CT scans. Yet, in practice, deep learning-based diffeomorphic methods struggle to capture large deformations between inspiratory and expiratory volumes, and therefore lack inverse consistency. Existing methods also fail to account for model uncertainty, which can be useful for improving performance. We propose an uncertainty-aware test-time adaptation framework for inverse consistent diffeomorphic lung registration. Our method uses Monte Carlo (MC) dropout to estimate spatial uncertainty that is used to improve model performance. We train and evaluate our method for inspiratory-to-expiratory CT registration on a large cohort of 675 subjects from the COPDGene study, achieving a higher Dice similarity coefficient (DSC) between the lung boundaries (0.966) compared to both VoxelMorph (0.953) and TransMorph (0.953). Our method demonstrates consistent improvements in the inverse registration direction as well with an overall DSC of 0.966, higher than VoxelMorph (0.958) and TransMorph (0.956). Paired t-tests indicate statistically significant improvements. 
\end{abstract}
\section{Introduction}
\label{sec:intro}
Deformable image registration (DIR) establishes a dense correspondence between two medical image volumes, playing a critical role in tasks such as disease phenotyping~\cite{bodduluri2017biomechanical} and surgical guidance~\cite{pelanis2021evaluation}. While DIR is widely used for image matching, its displacement fields lack guarantees of smoothness and invertibility, especially when handling large deformations. For instance, when registering chest computed tomography (CT) volumes at end-inspiration and end-expiration, large deformations could lead to inadequate displacements. This is further exacerbated in cases where inverse consistency between transformations is also required.

\begin{figure}[!t]
\centering
\includegraphics[scale = 1.35, center, trim={0.8cm 0.60cm 1.60cm 0.70cm}, clip]{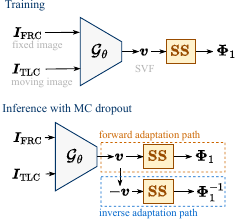}
\caption{Inverse consistent test-time adaptation framework. During training, the network $\mathcal{G}_{\theta}$ learns to predict SVF $\boldsymbol{v}$, from a fixed image $\boldsymbol{I}_\mathrm{F} = \boldsymbol{I}_\mathrm{FRC}$ and a moving image $\boldsymbol{I}_\mathrm{M} = \boldsymbol{I}_\mathrm{TLC}$. The SVF is then integrated using the scaling and squaring (SS) method to get the final displacement field $\mathbf{\Phi}_1$~\cite{arsigny2006log}. During inference, two different pathways (\textcolor{red}{forward} and \textcolor{blue}{inverse}) are defined for uncertainty estimation and adaptation, details of which are described below. The LDDMM framework allows for adaptation in inverse direction by simply negating the SVF ($-\boldsymbol{v}$) and then integrating it through SS.}
\label{overview}
\end{figure}

Large deformation diffeomorphic metric mapping (LDDMM) offers a framework for estimating transformations with theoretical guarantees of smoothness, differentiability, and invertibility, enabling consistent, one-to-one mappings even for images with large deformations~\cite{beg2005computing}. This one-to-one mapping also preserves topological consistency across images. While several iterative optimization methods have been developed to estimate these diffeomorphisms, they remain computationally intensive and time-consuming in practice~\cite{ashburner2007fast}. Recently, deep neural networks have been used to estimate diffeomorphic transformations that are faster and less computationally expensive at inference~\cite{mok2020fast,balakrishnan2019voxelmorph,chen2022transmorph}. These methods use convolutional neural networks (CNNs) or vision transformer backbones to directly predict a stationary velocity field (SVF), which is integrated to estimate the displacement field~\cite{mok2020fast}. While these methods utilize a theoretically informed framework (LDDMM) to estimate diffeomorphic transformations, they still lead to inadequate displacements in regions with large deformations which may not be readily invertible.

Most state-of-the-art registration methods also overlook model uncertainty, which is an important measure of confidence associated with predictions. We believe that incorporating uncertainty estimation could identify low and high confidence regions within the predicted displacement fields that could be used for rapid model adaptation at test time. We propose an uncertainty-aware test-time adaptation framework for inverse consistent diffeomorphic lung image registration that can be used to enhance model performance in both forward and inverse registration directions. Using Monte Carlo (MC) dropout, we generate spatial uncertainty maps at inference and use them to adapt and optimize the model at test-time~\cite{gal2016dropout}. Our framework demonstrates robust adaptation for bidirectional registration from total lung capacity (TLC) CT scans to functional residual capacity (FRC) scans and vice versa.

\section{Methods}
\label{sec:format}

\subsection{Variational Image Registration Framework}
Deformable image registration (DIR) is a highly ill-posed problem. We used the variational DIR framework that finds an optimal transformation $\mathbf{\Phi}: \Omega_F \rightarrow \Omega_M$ to align a fixed image $\boldsymbol{I}_\mathrm{F}: \Omega_F \subset \mathbb{R}^3 \rightarrow \mathbb{R}$ and a moving image $\boldsymbol{I}_\mathrm{M}: \Omega_M \subset \mathbb{R}^3 \rightarrow \mathbb{R}$, defined over the domains $\Omega_F$ and $\Omega_M$, respectively. The transformation $\mathbf{\Phi}$ deforms $\Omega_M$ to match $\Omega_F$, creating an alignment between $\boldsymbol{I}_\mathrm{F}$ and $\boldsymbol{I}_\mathrm{M} \circ \mathbf{\Phi}$, where $\boldsymbol{I}_\mathrm{M} \circ \mathbf{\Phi} = \boldsymbol{I}_\mathrm{D}$ is the warped moving image. The variational cost function $\mathcal{J}(\boldsymbol{I}_\mathrm{F}, \boldsymbol{I}_\mathrm{M}, \mathbf{\Phi})$ is typically composed of a data fidelity term $\mathcal{D}$ and a regularization term $\mathcal{R}$:    
\[
\mathcal{J}(\boldsymbol{I}_\mathrm{F}, \boldsymbol{I}_\mathrm{M}, \mathbf{\Phi}) = \underbrace{\mathcal{D}(\boldsymbol{I}_\mathrm{F}, \boldsymbol{I}_\mathrm{M} \circ \mathbf{\Phi})}_\text{image similarity~} + \underbrace{\lambda \, \mathcal{R}(\mathbf{\Phi})}_\text{regularization~},
\]
where $\mathcal{D}$ measures the similarity between $\boldsymbol{I}_\mathrm{F}$ and the deformed moving image $\boldsymbol{I}_{\mathrm{D}} = \boldsymbol{I}_\mathrm{M} \circ \mathbf{\Phi}$, and $\mathcal{R}(\mathbf{\Phi})$ enforces smoothness or other regularization properties on $\mathbf{\Phi}$. The scalar $\lambda > 0$ controls the balance between the image similarity and regularization terms.

Most deep learning registration networks typically parameterize this framework using a convolutional neural networks~\cite{balakrishnan2019voxelmorph, hering2021cnn} or vision transformers~\cite{chen2022transmorph} that learn to predict a dense displacement field $\boldsymbol{u}(\mathbf{x}) = \mathbf{\Phi}_{\boldsymbol{u}}(\mathbf{x}) - \mathbf{x}, \mathbf{x} \in \Omega_{F}$ from the fixed image space to the moving image space. Although this framework enables fast DIR, it lacks guarantees of smooth, invertible transformations, making the estimated large deformations potentially unreliable even with the smoothness constraints.

\begin{algorithm}[!t]
\scriptsize
    \caption{\textcolor{blue}{Uncertainty-Aware Test-Time Adaptation}}
    \label{alg:test_time_adaptation}
    \begin{algorithmic}[1]
        \State \textbf{\textcolor{blue}{Input:}} \textcolor{violet}{\( \boldsymbol{I}_{\mathrm{F}} \)}, \textcolor{violet}{\( \boldsymbol{I}_{\mathrm{M}} \)}, adaptation steps \textcolor{violet}{\( \mathcal{T} \)}, number of MC samples \textcolor{violet}{\( \mathcal{N} \)}, model \textcolor{violet}{\( \mathcal{G}_\theta(\boldsymbol{I}_{\mathrm{F}}, \boldsymbol{I}_{\mathrm{M}}) \)}.
        \State \textcolor{red}{\( \boldsymbol{v}_{\mathrm{MC}} \gets \emptyset \)}
        \For{\textcolor{teal}{sample = 1 to \( \mathcal{N} \)}}
            \State \( \mathbf{v}^{(i)} \gets \mathcal{G}_\theta(\boldsymbol{I}_{\mathrm{F}}, \boldsymbol{I}_{\mathrm{M}}) \) \Comment{\textcolor{gray}{MC Dropout}}
            \State \( \boldsymbol{v}_{\mathrm{MC}} \gets \boldsymbol{v}_{\mathrm{MC}} \cup \mathbf{v}^{(i)} \)
        \EndFor
        \State \( \bar{\boldsymbol{v}} \gets \frac{1}{\mathcal{N}} \sum_{i=1}^{\mathcal{N}} \boldsymbol{v}^{(i)} \) \Comment{\textcolor{gray}{Compute mean velocity field}}
        \State \( \sigma^2_{\boldsymbol{v}} \gets \frac{1}{\mathcal{N}} \sum_{i=1}^{\mathcal{N}} (\boldsymbol{v}^{(i)} - \bar{\boldsymbol{v}})^2 \) \Comment{\textcolor{gray}{Compute uncertainty map}}
       
        \For{\textcolor{teal}{step = 1 to \( \mathcal{T} \)}}
            \State \( \mathcal{L}_{\mathrm{total}} \gets \frac{1}{\sigma^2_{\boldsymbol{v}}}\mathcal{L}_{\mathrm{MSE}}(\boldsymbol{I}_\mathrm{F}, \boldsymbol{I}_\mathrm{M}) + \lambda \mathbf{\Gamma}(\mathrm{\phi}) \)
            \Comment{\textcolor{gray}{Compute uncertainty-weighted loss function}}
            
            \State Update \textcolor{violet}{\( \mathcal{G}_\theta \)} by minimizing \textcolor{red}{\( \mathcal{L}_{\mathrm{total}} \)}
        \EndFor
        \State \textbf{\textcolor{blue}{Output:}} Adapted model \textcolor{violet}{\( \mathcal{G}_\theta \)}
    \end{algorithmic}
\end{algorithm}

\subsection{Diffeomorphic Image Registration}
To ensure that the transformation estimated by our network is smooth and invertible, we used diffeomorphic deformations fields $\phi_t$ defined over time $t \in [0, 1]$~\cite{beg2005computing,mok2020fast}. The path of these fields can be generated by the velocity fields $\boldsymbol{v}_t$ as:
\[
\frac{d\mathbf{\Phi}_t}{dt} = \boldsymbol{v}_t(\mathbf{\Phi}_t) = \boldsymbol{v}_t \circ \mathbf{\Phi}_t, \quad \mathbf{\Phi}_0(\mathbf{x}) = \mathbf{x}.
\]
where $\boldsymbol{v}_t$ denotes the velocity at time $t$, and $\mathbf{\Phi}_0$ is the identity transform. Similar to~\cite{mok2020fast}, we predicted a SVF $\boldsymbol{v} = \mathcal{G}_{\theta}(\boldsymbol{I}_\mathrm{F}, \boldsymbol{I}_\mathrm{M})$, that was used to estimate $\mathbf{\Phi}_1$ by approximating the integration over $K$ time steps using the scaling and squaring method~\cite{arsigny2006log} (see Fig.~\ref{overview}). The diffeomorphic transformation $\mathbf{\Phi}_1$ was used to deform the moving image $\boldsymbol{I}_{\mathrm{M}}$ into the fixed image space $\Omega_F$ to get the deformed image $\boldsymbol{I}_{\mathrm{D}} = \boldsymbol{I}_{\mathrm{M}} \circ \mathbf{\Phi}_1$. The image similarity loss was defined as the mean-squared error between the fixed image $\boldsymbol{I}_\mathrm{F}$ and the deformed moving image $\boldsymbol{I}_{\mathrm{D}} = \boldsymbol{I}_\mathrm{M} \circ \mathbf{\Phi}_1$:
\[
\mathcal{L}_{\text{MSE}}(\boldsymbol{I}_\mathrm{F}, \boldsymbol{I}_\mathrm{M}) = \frac{1}{|\Omega_F|} \int_{\Omega_F} \left( \boldsymbol{I}_\mathrm{F}(\mathbf{x}) - \boldsymbol{I}_{\mathrm{D}}(\mathbf{x}) \right)^2 \, d\mathbf{x},
\]
where $|\Omega_F|$ denoted the volume of the domain $\Omega_F$. To encourage smooth and plausible deformations, we used a bending energy regularization loss to penalize large second-order spatial derivatives in the velocity field $\boldsymbol{v}$. The bending energy regularization loss was defined as:

\begin{equation}
\mathbf{\Gamma}(\mathbf{\Phi}_1) = \frac{1}{|\Omega_F|} \int_{0}^{U} \int_{0}^{V} \int_{0}^{W} \left\| \nabla^2 \mathbf{\Phi}_1 \right\|_F^2 \, d\mathbf{u}d\mathbf{v}d\mathbf{w},
\end{equation}
where $U$, $V$, $W$ denoted the 3D spatial dimensions of the dense displacement field $\mathbf{\Phi}_1$. The combined loss function was expressed as a linear combination of image similarity and displacement smoothness:
\[
\mathcal{L}_{\text{total}} = \mathcal{L}_{\text{MSE}} + \lambda \mathbf{\Gamma}(\mathbf{\Phi}).
\]
For training our model $\mathcal{G}_{\theta}$, we used the FRC image as the fixed image $\boldsymbol{I}_\mathrm{F} = \boldsymbol{I}_\mathrm{FRC}$ and the TLC image as the moving image $\boldsymbol{I}_\mathrm{M} = \boldsymbol{I}_\mathrm{TLC}$ see Fig.~\ref{overview}.

\begin{figure}[!t]
\centering
\includegraphics[scale = 0.07]{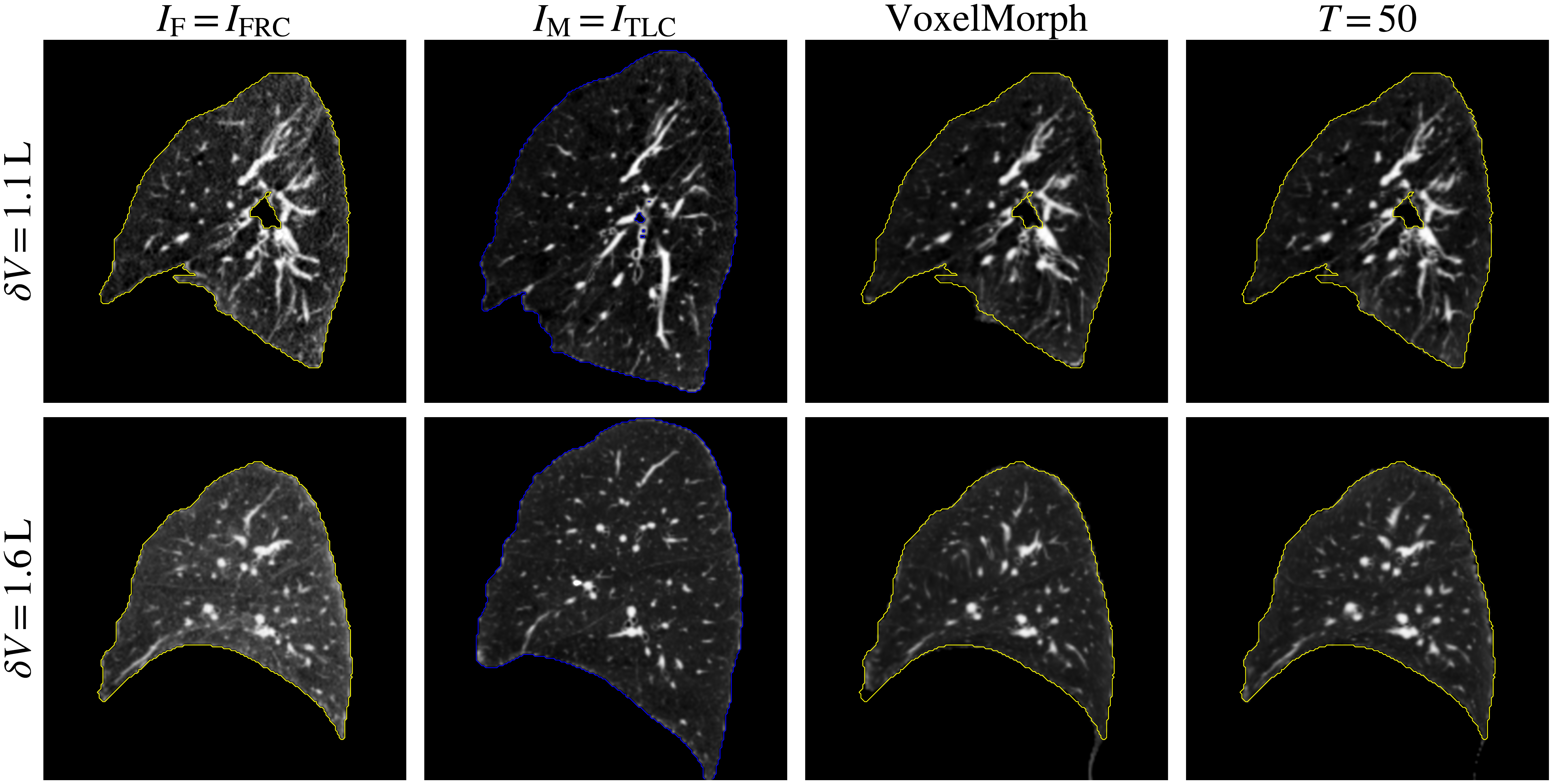}
\caption{Qualitative visualization of two large deformation \textcolor{red}{forward} registration cases with $\delta V = 1.1$ L and $\delta V = 1.6$ L. The fixed image outlines, shown in yellow, are overlaid on the deformed images for references. Change in volume was defined in liters (L).}
\label{tlc_to_frc}
\end{figure}

\subsection{Network Architecture}
We parameterized our diffeormorphic image registration model $\mathcal{G}_{\theta}(\boldsymbol{I}_{\mathrm{F}}, \boldsymbol{I}_{\mathrm{M}})$ using a 3D UNet-like CNN architecture augmented with residual connections for improved model performance~\cite{kerfoot2019left}. The input to the network were single-channel, 3D fixed $\boldsymbol{I}_{\mathrm{F}} \in \mathbb{R}^{1 \mathrm{\times H \times W \times D}}$ and moving $\boldsymbol{I}_{\mathrm{M}} \in \mathbb{R}^{1 \times \mathrm{H \times W \times D}}$ image volumes. The encoder and decoder backbones respectively had 6 downsampling and upsampling convolutional blocks that were connected through skip connections~\cite{kerfoot2019left}. At the bottleneck, we used a single residual block for improved knowledge transfer across the encoder and decoder pathways. The output of the network was a three-channel SVF denoted as $\boldsymbol{v} \in \mathbb{R}^{3 \mathrm{\times H \times W \times D}}$, as shown in Fig.~\ref{overview}. For integrating the SVF $\boldsymbol{v}$, we used a differentiable scaling and squaring (SS) layer to get the final displacement field  $\mathbf{\Phi}_1$ (see Fig.~\ref{overview}), which was applied to the moving image~\cite{arsigny2006log}.

\subsection{Uncertainty-Aware Test-Time Adaptation}
We propose an uncertainty-aware test-time adaptation framework for refining the registration of inspiratory and expiratory chest CT scans in both directions (see Fig.~\ref{overview} and Algorithm~\ref{alg:test_time_adaptation}). At inference, we first estimated the epistemic (or model) uncertainty using Monte Carlo (MC) dropout~\cite{gal2016dropout}. This was done by performing $\mathcal{N}$ forward passes through $\mathcal{G}_\theta$ to get a set of MC SVFs $\boldsymbol{v}_{\mathrm{MC}} = \{\boldsymbol{v}^{(i)}\}_{i=1}^\mathcal{N}$ (see Algorithm~\ref{alg:test_time_adaptation}). The mean and variance of these velocity fields provided a Bayesian estimate of the velocity field $\bar{\boldsymbol{v}}$ and model uncertainty, respectively defined as:
\[
\bar{\boldsymbol{v}} = \frac{1}{\mathcal{N}} \sum_{i=1}^\mathcal{N} \boldsymbol{v}^{(i)}, \quad \sigma^2_{\boldsymbol{v}} = \frac{1}{\mathcal{N}} \sum_{i=1}^\mathcal{N} (\boldsymbol{v}^{(i)} - \bar{\boldsymbol{v}})^2.
\]
We used the uncertainty map $\sigma^2_{\boldsymbol{v}}$ to identify the regions where the model was less confident, and applied a weighted correction based on $1 / \sigma^2_{\boldsymbol{v}}$ to encourage model performance under confident regions. The adaptation for each sample was conducted for $\mathcal{T}$ steps using the uncertainty weighted loss function shown in Algorithm~\ref{alg:test_time_adaptation}. Although the model was trained for TLC (moving) to FRC (fixed) registration, adaptation method was applied to refine both TLC to FRC and FRC to TLC registrations. The latter was conducted by simply negating the SVF and integrating it through the SS layer (see Fig.~\ref{overview}).

\begin{figure}[!t]
\centering
\includegraphics[scale = 0.07]{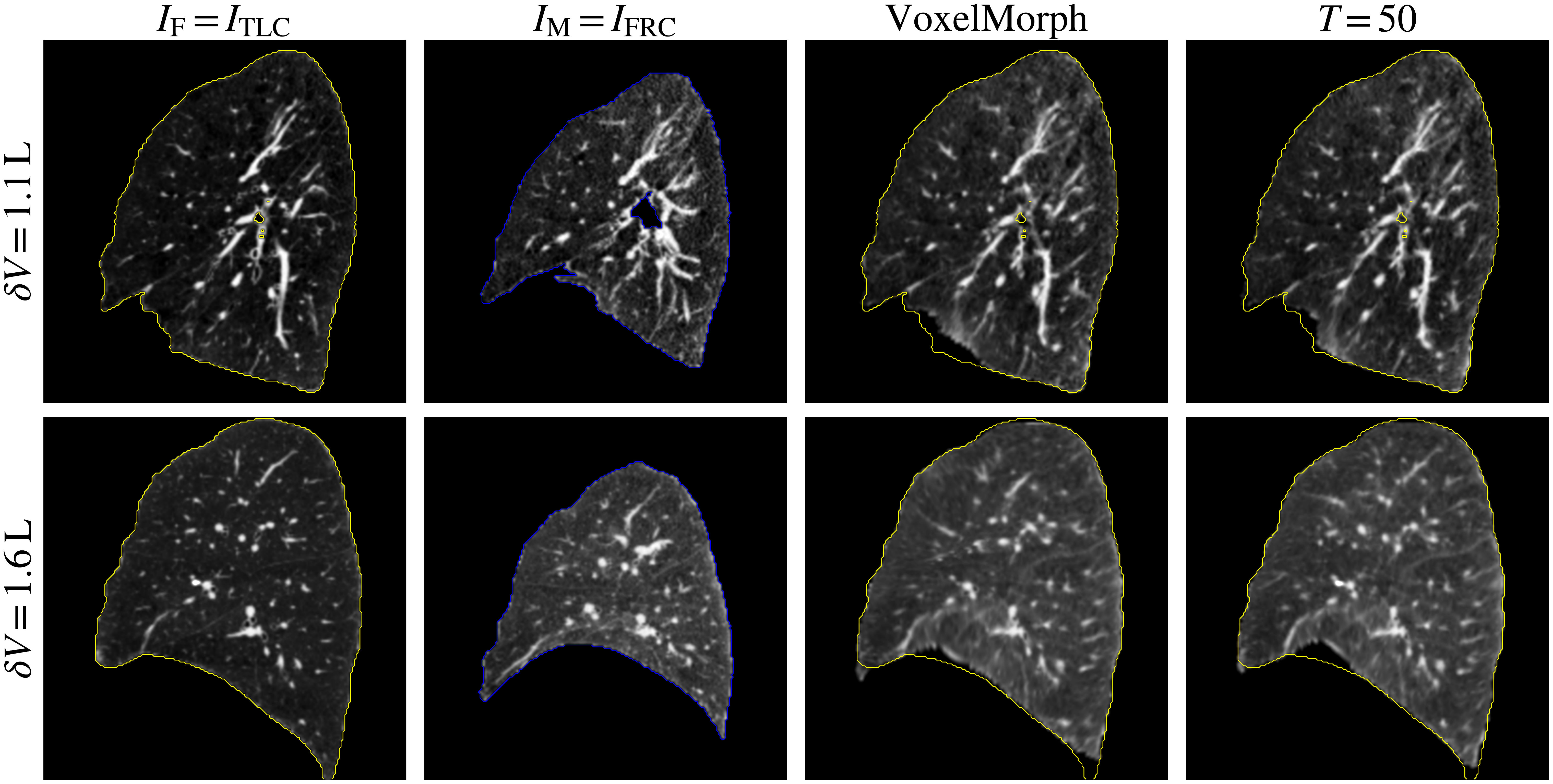}
\caption{Visualization of \textcolor{blue}{inverse} image registration with large $\delta V = 1.1$ L and $\delta V = 1.6$ L. The inverse transformation $\mathbf{\Phi}_{1}^{-1}$ was obtained by the integration of negative SVF $-\boldsymbol{v}$.}
\label{frc_to_tlc}
\end{figure}

\begin{table*}[!t]
\centering
\scriptsize
\setlength{\tabcolsep}{4pt} 
\caption{Quantitative evaluation using lung mask overlap assessed through DSC and ASSD in mm. We also report percent folding within the lung region, quantified by $\lvert \boldsymbol{J} \rvert_{\mathrm{\Omega}} < 0$ ($\%$). We used paired Student's t-test to compare VoxelMorph and TransMorph with different test-time adaptation models at $\mathcal{T} = \{10, 30, 50 \}$.}
\begin{tabularx}{0.725\textwidth}{l | c c c | c c c}
    \toprule
    & \multicolumn{3}{c|}{\textbf{\textcolor{red}{Forward}}: $\boldsymbol{I}_{\mathrm{F}} = \boldsymbol{I}_{\mathrm{FRC}},\: \boldsymbol{I}_{\mathrm{D}} = \boldsymbol{I}_{\mathrm{TLC}} \circ \mathbf{\Phi}_1$} & \multicolumn{3}{c}{\textbf{\textcolor{blue}{Inverse}}: $\boldsymbol{I}_{\mathrm{F}} = \boldsymbol{I}_{\mathrm{TLC}},\: \boldsymbol{I}_{\mathrm{D}} = \boldsymbol{I}_{\mathrm{FRC}} \circ \mathbf{\Phi}_{1}^{-1}$} \\
    \cmidrule(lr){2-4} \cmidrule(lr){5-7}
    & \textbf{DSC}\textsubscript{$\uparrow$} & \textbf{ASSD} (mm)\textsubscript{$\downarrow$} & $\lvert \boldsymbol{J} \rvert_{\mathrm{FRC}} < 0$ (\%)\textsubscript{$\downarrow$} 
    & \textbf{DSC}\textsubscript{$\uparrow$} & \textbf{ASSD} (mm)\textsubscript{$\downarrow$} & $\lvert \boldsymbol{J} \rvert_{\mathrm{TLC}} < 0$ (\%)\textsubscript{$\downarrow$} \\
    \midrule
    VoxelMorph & 0.953 $\pm$ {0.026}$^{\star}$ & 1.37 $\pm$ {0.93}$^{\star}$ & $ < 1\%$ 
               & 0.958 $\pm$ {0.025}$^{\star}$ & 1.26 $\pm$ {0.81}$^{\star}$ & $ < 1\%$ \\
    TransMorph & 0.953 $\pm$ {0.023}$^{\star}$ & 1.28 $\pm$ {0.80}$^{\star}$ & $ < 1\%$ 
               & 0.956 $\pm$ {0.017}$^{\star}$ & 1.23 $\pm$ {0.59}$^{\star}$ & $ < 1\%$ \\
    \midrule
    $\mathcal{T} = 10$   & 0.961 $\pm$ {0.021} & 1.18 $\pm$ {0.81} & $ < 1\%$ 
                                    & 0.965 $\pm$ {0.022} & 1.06 $\pm$ {0.72} & $ < 1\%$ \\
    $\mathcal{T} = 30$   & 0.965 $\pm$ {0.020} & 1.11 $\pm$ {0.77} & $ < 1\%$ 
                                    & 0.966 $\pm$ {0.023} & 1.03 $\pm$ {0.72} & $ < 1\%$ \\
    $\mathcal{T} = 50$   & 0.966 $\pm$ {0.020} & 1.09 $\pm$ {0.77} & 1.8$\%$ 
                                    & 0.965 $\pm$ 0.025 & 1.04 $\pm$ 0.75 & $ < 1\%$ \\
    \bottomrule
\end{tabularx}
\label{tab:quant_def}
\end{table*}

\section{Experimental Design}

\subsection{Dataset and Image Preprocessing}
We trained and evaluated our models on 675 TLC and FRC CT volume pairs from the Genetic Epidemiology of COPD (COPDGene) study~\cite{regan2011genetic}. We randomly sampled 100 chest CT images across different COPD severity stages, defined from I (mild) to IV (severe), by the Global Initiative for Chronic Obstructive Lung Disease (GOLD)~\cite{agusti2023global}. In addition to that, our analysis cohort also included data from 100 asymptomatic smokers (GOLD 0), 100 individuals with preserved ratio and impaired spirometry (PRISm), and 75 never-smoking normals~\cite{regan2011genetic}. We split our data into non-overlapping training ($50\%$), validation ($25\%$), and held-out test ($25\%$) sets. Before training our models, we isotropically resampled our images to $\mathrm{1.5} \times \mathrm{1.5} \times \mathrm{1.5}$ mm\textsuperscript{3}. To reduce overall computational complexity, we trained our models using cropped left and right lungs. Before training, we clipped the chest CT images intensities between -1024 Hounsfield units (HU) and 1024, and rescaled them to the interval $[-1, 1]$.

\subsection{Evaluation Metrics and Comparison Methods}
We evaluated our model by quantifying the overlap between the fixed and deformed image masks using the Dice similarity coefficient (DSC) and average symmetric surface distance (ASSD)~\cite{heimann2009comparison}. We also calculated the percentage of Jacobian determinant below zero, $\lvert \boldsymbol{J} \rvert_{\mathrm{\Omega}} < 0$ ($\%$) to evaluate folding. We compared our methods with the baseline diffeomorphic variant of the well-known VoxelMorph and TransMorph frameworks~\cite{balakrishnan2019voxelmorph, chen2022transmorph}.

\begin{figure}[!h]
    \centering
    \includegraphics[scale = 0.12]{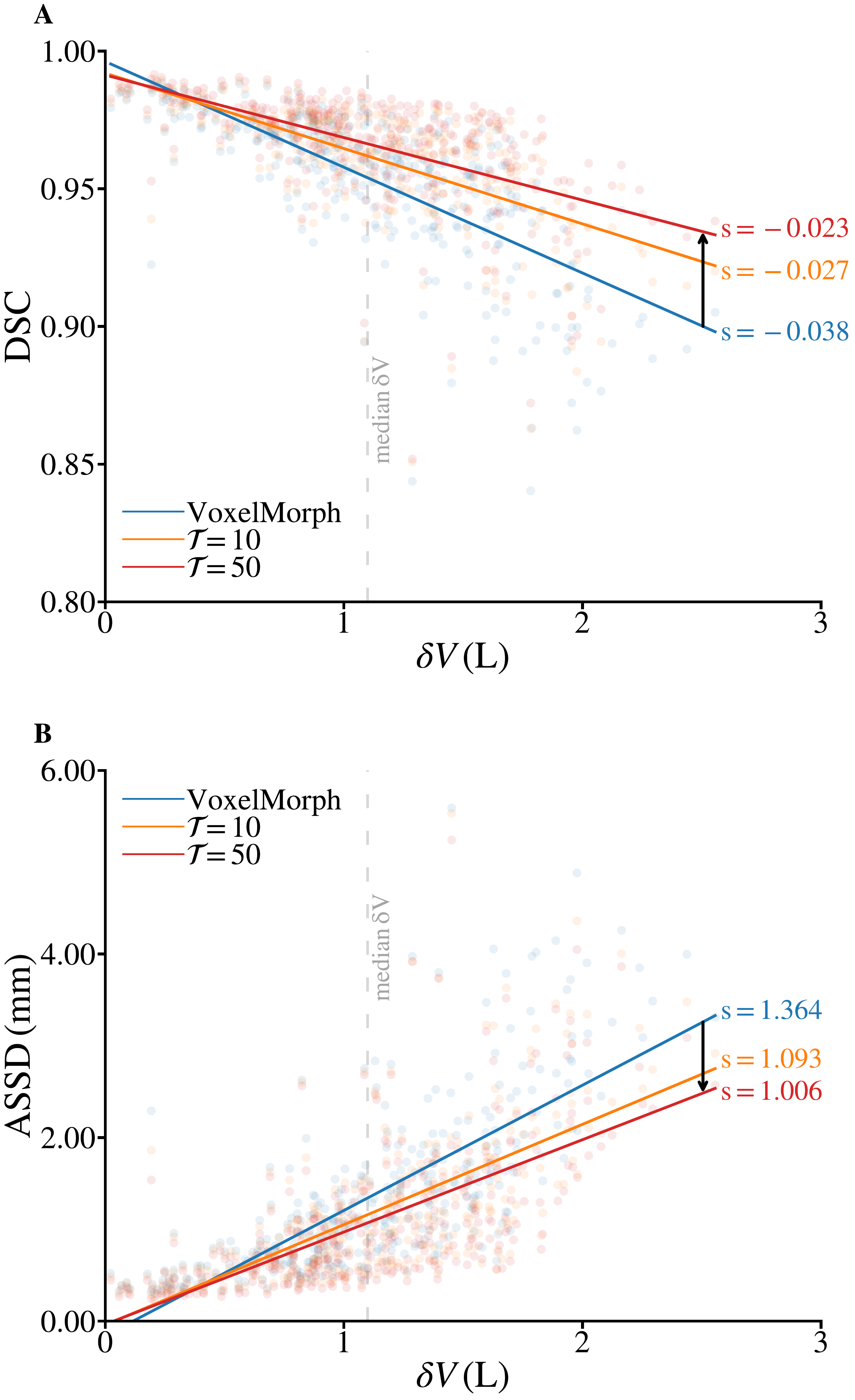}
    \caption{Change in lung mask overlap metrics [(A) DSC and (B) ASSD] for TLC to FRC (\textcolor{red}{forward}) with increasing $\delta V$s. Regression lines for each method and their slopes indicate improved model performance with increasing test-time adaptation steps $\mathcal{T}$.}
    \label{fig:performance_analysis}
\end{figure}

\subsection{Implementation Details}
All models were implemented using open source frameworks PyTorch and MONAI~\cite{cardoso2022monai}. During training and test-time adaptation, we used a learning rate of 0.0002 and a $\lambda = 0.2$. The SS method used $K = 10$ timesteps for integrating the SVFs. For estimating model uncertainty we used $\mathcal{N} = 20$ forward passes, and test-time adaptation was conducted for different number of steps $\mathcal{T} = \{10, 30, 50\}$. All models were training using a single NVIDIA A6000 GPU with 48 GB memory.

\section{Results}
We present qualitative visualization of the test-time adaptation for TLC to FRC and FRC to TLC image registration in Figs.~\ref{tlc_to_frc} and~\ref{frc_to_tlc}, respectively. Quantitative comparison with diffeomorphic VoxelMorph and TransMorph is presented in Table~\ref{tab:quant_def}. All adapted models achieved significantly higher performance ($P < 0.001$) compared to VoxelMorph and TransMorph. DSCs and ASSDs (mm) are also studied against the increasing change in volume $\delta V$ (L) (see Fig.~\ref{fig:performance_analysis}). As shown in Fig.~\ref{fig:performance_analysis}A, the regression lines for VoxelMorph became less steep due an increase in DSCs with $\delta V$. A similar improvement trend was observed for ASSD (mm) slopes shown in Fig.~\ref{fig:performance_analysis}B.

\section{Discussion}
We developed an uncertainty-aware test-time adaptation framework for inverse-consistent, diffeomorphic lung image registration. Spatial uncertainty maps, estimated through MC dropout were used to inform test-time adaptation, allowing the model to dynamically adjust its predictions for improved registration accuracy in both forward (TLC to FRC) and inverse (FRC to TLC) directions. Our method was evaluated on large cohort of COPD patients with varying degrees of disease severity, which manifested in a cohort with a broad range of volume changes. The median volume change per lung was 1.1 L which made this cohort well-suited for evaluating inverse consistent image registration under large deformations. 

Although our models were trained for TLC to FRC registration, our method allowed for adaptation from FRC to TLC as well. We acknowledge that our method requires a certain number of test-time adaptation steps which may increase the overall time for registration. However, adaptation for each sample still took less than a minute even for 50 steps. Another important capability offered by our framework is that it allows for anatomical regularization at test-time. Future work could extend this framework by exploring alternative uncertainty estimation techniques, such as Bayesian neural networks, or heteroscedastic uncertainty estimation using log-likelihoods. 

\section{Compliance with Ethical Standards}
This research study was conducted retrospectively using anonymized human subject data made publicly available by COPDGene. Written consent was provided by all subjects, and the protocols were approved by the Institutional Review Boards (IRBs) of each participating study center.

\section{Acknowledgments}
\label{sec:acknowledgments}
This work was supported in part by the grants R01HL151421 and K01HL163249. COPDGene was supported by NHLBI grants U01 HL089897 and U01 HL089856 and by NIH contract 75N92023D00011. The COPDGene study (NCT00608-764) has also been supported by the COPD Foundation through contributions made to an Industry Advisory Committee that has included AstraZeneca, Bayer Pharmaceuticals, Boehringer-Ingelheim, Genentech, GlaxoSmithKline, Novartis, Pfizer, and Sunovion.

\bibliographystyle{IEEEbib}
\bibliography{refs}

\end{document}